\documentclass{jps-cp}
\usepackage{txfonts} 
\usepackage[allow-number-unit-breaks]{siunitx}

\title{Josephson Parametric Amplifier in Axion Experiments}

\author{Jinmyeong \textsc{Kim}$^{2,1}$, Boris I. \textsc{Ivanov}$^{1}$, Çağlar \textsc{Kutlu}$^{2,1}$, Seongtae \textsc{Park}$^{1}$, Arjan F. \textsc{van Loo}$^{3,4}$, Yasunobu \textsc{Nakamura}$^{3,4}$, Sergey V. \textsc{Uchaikin}$^{1}$, Seonjeong \textsc{Oh}$^{1}$, Violeta \textsc{Gkika}$^{1}$, Andrei \textsc{Matlashov}$^{1}$, Woohyun \textsc{Chung}$^{1}$, and Yannis K. \textsc{Semertzidis}$^{1,2}$}

\inst{$^{1}$Center for Axion and Precision Physics Research, IBS, Daejeon 34051, Republic of Korea \\
$^{2}$Department of Physics, KAIST, Daejeon 34141, Republic of Korea \\
$^{3}$RIKEN Center for Quantum Computing (RQC), Wako, Saitama 351–0198, Japan \\
$^{4}$Department of Applied Physics, Graduate School of Engineering, The University of Tokyo, Bunkyo-ku, Tokyo 113-8656, Japan}

\email{ambritjm@kaist.ac.kr}

\recdate{July 15, 2022}

\abst{
The axion is a hypothetical particle, a promising candidate for dark matter, and a solution to the strong CP problem. Axion haloscope search experiments deal with a signal power comparable to noise uncertainty at millikelvin temperature. We use a flux-driven Josephson parametric amplifier~(JPA) with the aim of approaching a noise level near the theoretically allowed limit of half quanta. In our measurements to characterize the JPA we have found the added noise to the system with a JPA as the first-stage amplifier to be lower than \SI{110}{\milli \kelvin} at the frequencies from \SI{0.938}{\giga\hertz} to \SI{0.963}{\giga\hertz}.}
\kword{cryogenic noise source, superconducting circuit readout, amplifier, Josephson parametric amplifier, microwave amplifier}
\begin{document}
\maketitle
\section{Introduction}
The axion is a hypothetical particle proposed by Pecci and Quin to resolve the strong CP problem\cite{axion}. 
It is a strong candidate for dark matter, which is expected to cover about 85 \verb|%| of all matter in the universe\cite{cosmologyofaxion,cosmologyboundofaxion,notsoharmaxion}. The axion haloscope experiments search for axions by observing the microwave photons converted from axions in a magnetic field. Axion haloscope experiments use a cavity placed in a magnetic field in order to capture photons produced by axions penetrating the cavity\cite{axion}. An antenna coupled to the cavity measures the electrical signal power of the order of $10^{-22}$--\SI{e-23 }{\watt}. To scan over the unknown mass and coupling constant of an axion, the scan rate, which is inversely proportional to the system noise, is very important. Experiments with commercially available cryogenic amplifiers such as high-electron-mobility-transistor (HEMT) based amplifiers have a system noise temperature in the range of \SI{1}{\kelvin}--\SI{5}{\kelvin}, depending on the frequency. Experiments with Josephson parametric Amplifiers (JPA) can run at a noise level close to the quantum limit. Noise-temperature measurements of a JPA below \SI{120}{\milli\kelvin} have been achieved\cite{caglar, cas, JPACirc, Kerr, snail}. Several axion haloscope experiments have implemented JPAs\cite{JinSu22, admx, bru, Cres}. In this paper, we show the characterization of a prototype JPA chain which adds less than \SI{110}{\milli \kelvin} to the total noise temperature of the setup.

\section{JPA}
\subsection{Design and Theory}
\begin{figure}[ht]
\includegraphics[scale=0.4]{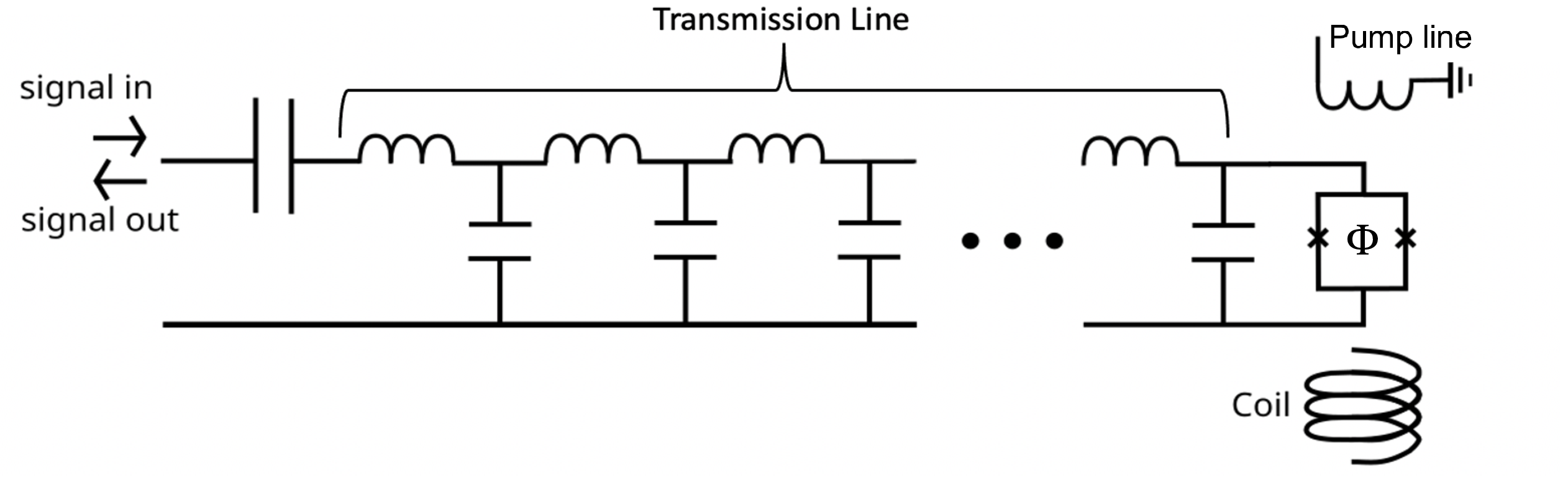}
\centering
\caption{Circuit representation of a JPA\cite{JPACirc}. Flux $\Phi$ through the loop is controlled by an inductively coupled coil and external pump line.}
\label{JPA Descript}
\end{figure}

Among the various JPA designs, we use a flux-driven JPA \cite{JPACirc} with a quarter-wavelength resonator terminated by a superconducting quantum interference device (SQUID), as shown in Fig.~\ref{JPA Descript}. The interdigitated capacitor, the coplanar-waveguide transmission line, and the coplanar-waveguide pump line are all made from niobium. The SQUID has an inner loop diameter of about \SI{16}{\micro\meter} and has two Josephson junctions with an identical critical current of about 2~$\mu$A. Both junctions are made with Al/AlO$_x$/Al technology. When the DC flux passing through the loop is fixed, a  quarter-wave resonance is formed by the capacitor, transmission line, and the effective inductance of the SQUID. The effective inductance is
\begin{equation}
    L_{eff} = \frac{\Phi_0}{2\pi I_c}\frac{1}{\cos{\pi\frac{\Phi}{\Phi_0}}},
\label{equ}
\end{equation}
where $\Phi_0$ is the flux quantum and $I_c$ is the critical current.
The pump line couples an AC flux to the loop. When the AC flux frequency is twice the resonance frequency, the JPA parametrically amplifies the signal. When an input signal mixes with a pumping signal, one more tone appears with the amplified signal at the output.

\subsection{RF-chain Schematic}
\begin{figure}[ht]
\includegraphics[scale=0.23]{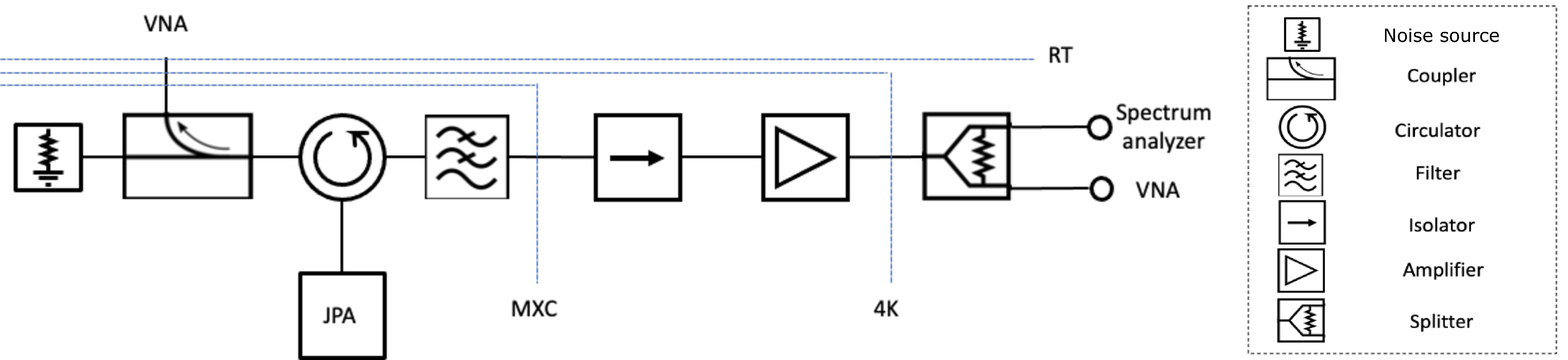}
\centering
\caption{Simplified model of the radio-frequency (RF) chain schematic. The directional coupler has a 20~dB attenuation to the signal line. The mixing-Chamber (MXC) is maintained at temperatures in the range of \SI{18}{\milli\kelvin} - \SI{50}{\milli\kelvin}. The gain and power spectrum are measured using the vector network analyzer~(VNA) and the spectrum analyzer~(SA) respectively.}
\label{JPA schem}
\end{figure}

The simplified RF chain is shown in Fig.~\ref{JPA schem}. The directional coupler introduces power to the chain from the 50~$\Omega$ terminated noise source (NS) through the input port and the signal from the VNA through the coupling port. The coupling line from the VNA has a total of 110~dB attenuation distributed across temperature stages to prevent gain saturation of the JPA and any noise from higher temperature stages. Signals passing the JPA, filter, and isolators are amplified by a commercial HEMT amplifier. We installed isolators and filters to reduce the noise to the JPA caused by backaction from the HEMT. Superconducting NbTi coaxial cables are used between NS, directional coupler, and other temperature stages to reduce heat flow and cable losses. At frequencies away from the resonance frequency, the JPA behaves like a perfect resonator with high out-of-band reﬂection at the input capacitor.

\subsection{Three-JPA PCB}
\begin{figure}[ht]
\includegraphics[scale=0.15]{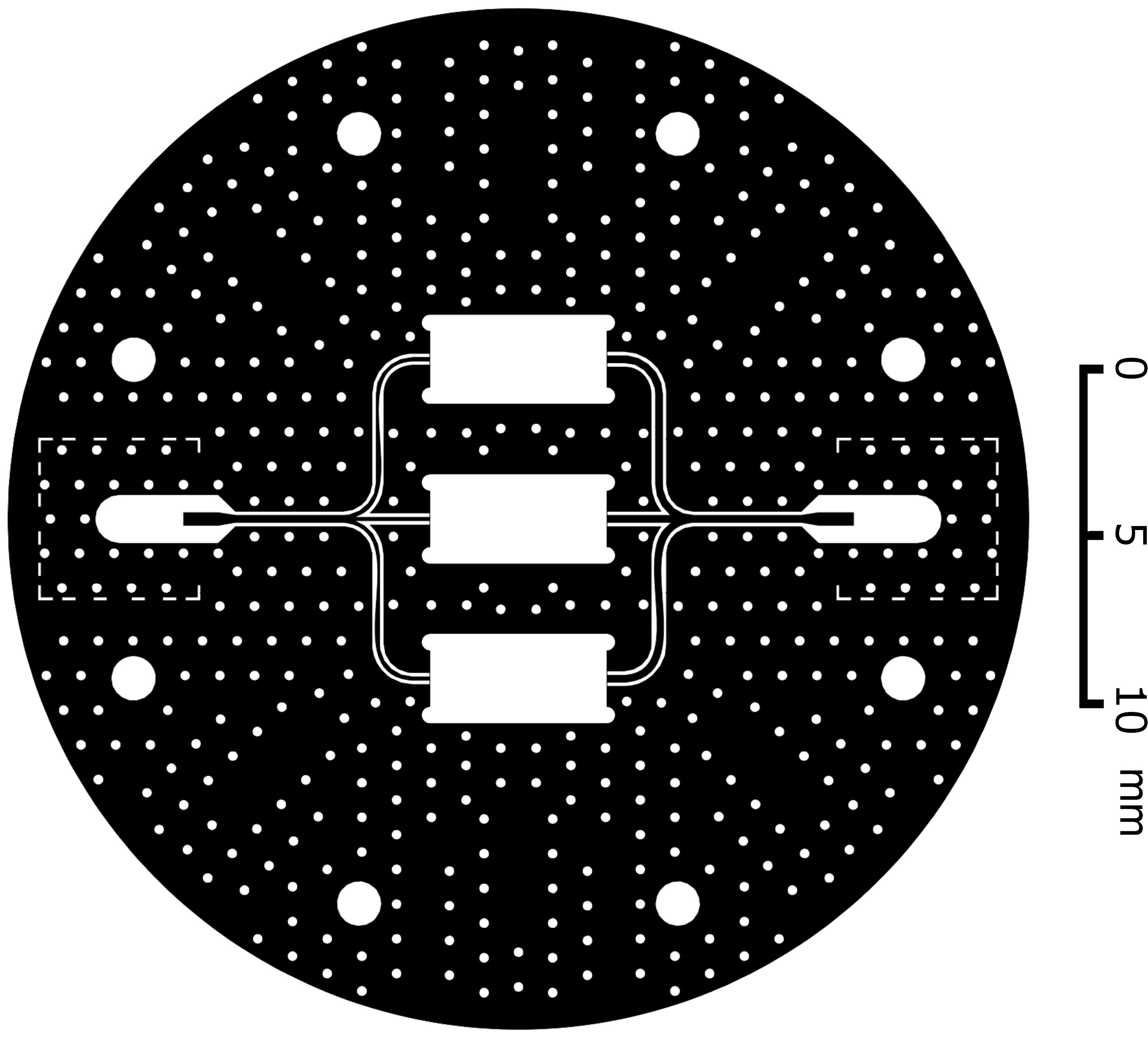}
\centering
\caption{Top view of the PCB design for the three parallelly connected JPA chips with a common signal line and separate pumping lines. Black area corresponds to gold-platted copper.}
\label{goryImage}
\end{figure}

We designed a printed circuit board (PCB) that has a three-way split signal input--output line such that three JPAs can be connected to a single signal input, see Fig.~\ref{goryImage}. The input and pump coplanar-waveguide transmission lines are split into three lines each split line go to a JPA. The length of the splitted line is about 5 mm and much smaller than the operating signal wavelength and the impedance mismatch is not significant.
Sub-Miniature Push-on~(SMP) connectors are placed on the high-frequency laminate for a better microwave property. A superconducting-wire coil is wound around the holder to provide a DC magnetic flux bias. The PCB and coil are inserted to a magnetic shield to reduce the effect of external magnetic fields.

\subsection{JPA-off measurements}
\begin{figure}[ht]
\includegraphics[scale=0.28]{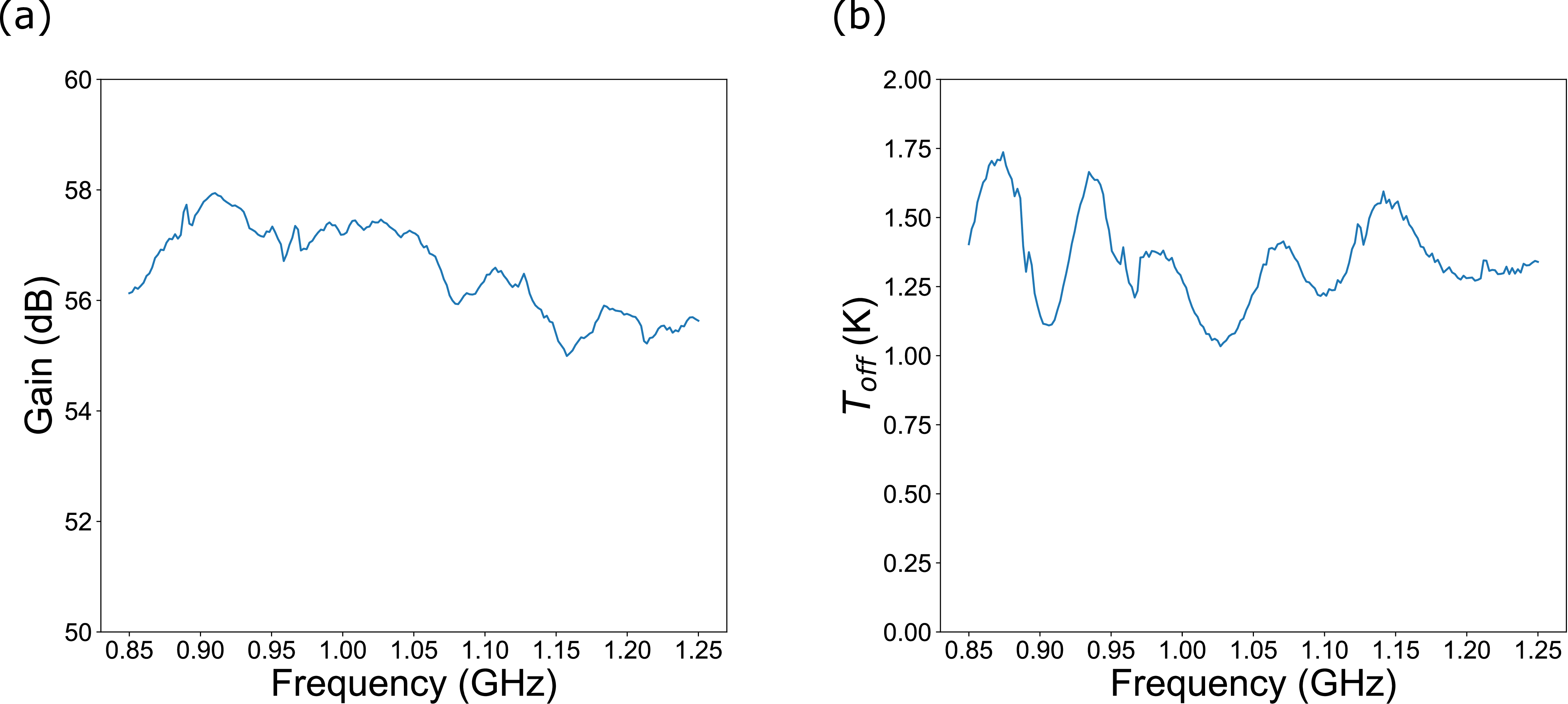}
\centering
\caption{(a)~JPA-\textit{off} gain and (b)~noise measurements using the NS; The gain and loss are mostly determined by the HEMT charecteristics, impedance mismatches, losses, and reflection due to components such as circulators, filters, isolators and cables.}
\label{JPA OFF}
\end{figure}

A JPA is \emph{off} when the resonance frequency is tuned away from the signal frequency by the DC flux and when the pump signal is off. When a JPA is \emph{off}, we estimate the noise and gain of the chain by measuring the power spectrum at the NS temperatures from \SI{0.1}{\kelvin} to \SI{0.4}{\kelvin}. The Y-factor method with reference to half quanta noise\cite{NS} was used to derive the noise temperature of the JPA \emph{off}. The gain and noise temperature of the RF chain with the JPA \emph{off} are plotted in Fig.~\ref{JPA OFF}.

\subsection{Flux sweep}
\begin{figure}[ht]
\includegraphics[scale=0.3]{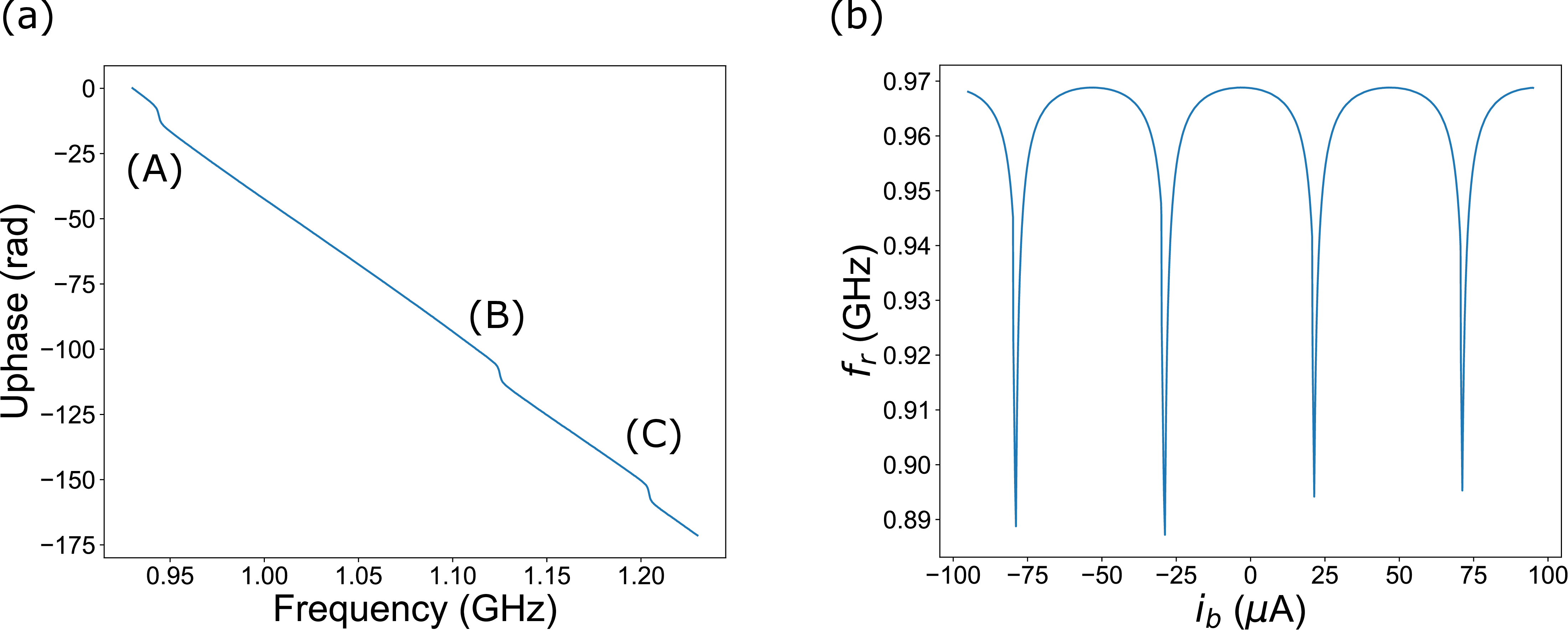}
\centering
\caption{JPA resonance frequency dependence on the flux bias. (a) Unwrapped phase measured over frequency. The three phase shifts correspond to the resonance frequencies of the three parallelly connected JPAs labeled A, B, and C. (b) Resonance frequency of JPA-A as a function of the flux coil current $i_b$.}
\label{res}
\end{figure}

The dependence of the JPA resonance frequency on the DC flux bias is shown in Fig.~\ref{res}. Here the phase shifts by the JPA resonance at frequencies \SI{0.944}{\giga\hertz}, \SI{1.125}{\giga\hertz}, and \SI{1.205}{\giga\hertz} are shown. For this paper, the JPA~(JPA-A) with the resonance at \SI{0.944}{\giga\hertz} is characterized. The quality factor of JPA-A is about 700. We tune the JPA resonance within a frequency range by controlling the flux coil current as in Fig.~\ref{res}(b). This JPA has a period of \SI{50}{\micro\ampere} shown in Fig.~\ref{res}(b). In the frequency range below 940 MHz the JPA has two stable points with diﬀerent frequencies. This results in a hysteric behavior in the ﬂux response\cite{hyst}.

\subsection{Gain map}
\begin{figure}[ht]
\includegraphics[scale=0.27]{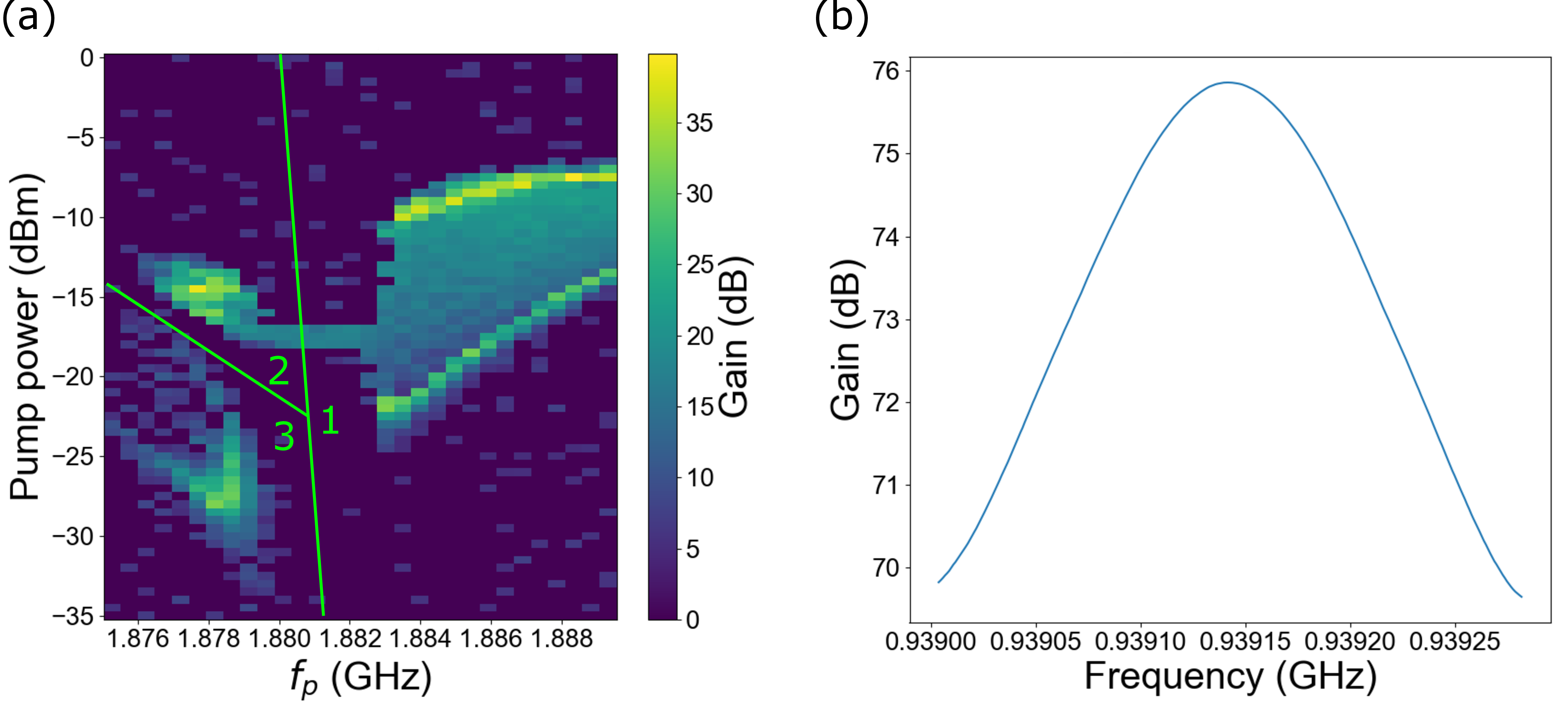}
\centering
\caption{(a) JPA gain map over pump power and pump frequency for a fixed flux bias current ($i_b$ = \SI{-27.14}{\micro\ampere}). The gain of the JPA-A is measured at a \SI{1}{\kilo\hertz} offset from $f_{p/2}$ to avoid measuring in the degenerate mode. (b) Total gain of the RF chain for a selected point in (a) ($P_p=$\SI{-29}{dBm}, $f_p=$\SI{1.878}{\giga\hertz}). A Savitzky-Golay filter with the window size of 100~kHz and the polynomial order of 3 is used.}
\label{gainmap}
\end{figure}

We scan over pump power ($P_p$), pump frequency ($f_p$), and flux coil current ($i_b$) to characterize the gain of JPA-A as shown in Fig.~\ref{gainmap}. There are various factors limiting these parameters for the axion experiment.  Increasing the gain of the JPA over \SI{21}{\decibel} reduces the bandwidth, which may reduce the practical scanning rate in axion search experiments by increasing the number of the tuning steps needed to cover the same frequency range. Three contour regions are shown in Fig.~\ref{gainmap} (a).As our measurement shows, Region 1 is not applicable because the bandwidth is half of that in Region 3. Region 2 is not used because the JPA is working not in the three-wave-mixing mode, but in a higher frequency mode with nonlinear behaviour. From Region 3 a set of points with gain between \SI{18}{\decibel} to \SI{21}{\decibel} is selected. Points with the lowest pump power at certain flux bias are further selected as working points.

\subsection{JPA on. Noise and Gain measurements}
\begin{figure}[ht]
\includegraphics[scale=0.3]{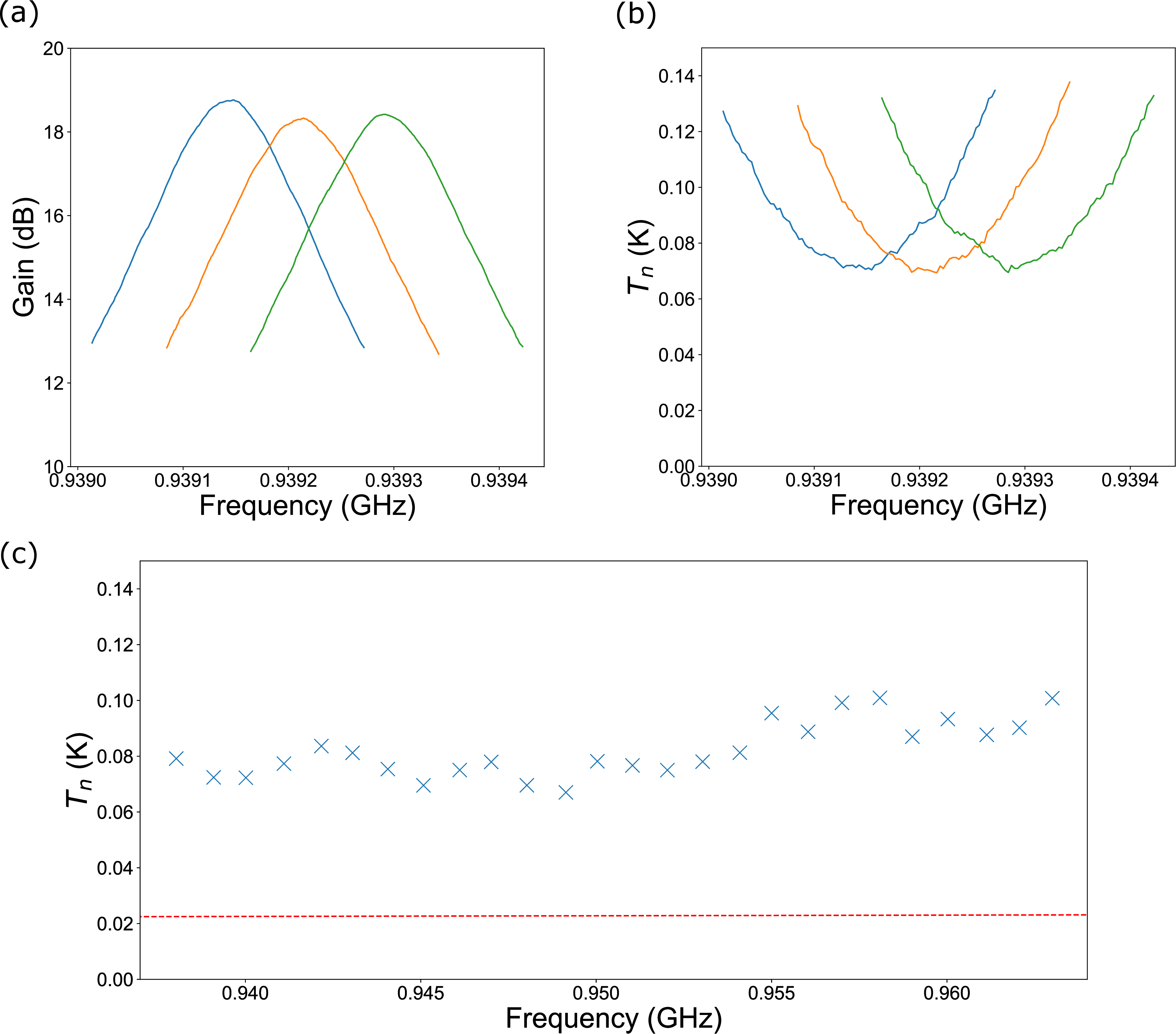}
\centering
\caption{Gain and noise for three selected working points of JPA-A are each shown in (a) and (b). To smooth the gain curve we use Savitzky–Golay window with a width of \SI{30}{\kilo\hertz} and polynomial order of 3. The minimum noise temperature of 25 different working points of JPA-A with \SI{1}{\mega\hertz} spacing is shown in (c). Standard quantum limit for added noise of half quanta ($hf/2k_B$) is show in (c) as a red dashed line.}
\label{gory}
\end{figure}

A JPA is described as being \emph{on} when the parameters are tuned to a working point. We take the power spectrum from the NS as a signal with the well known power ($P_{ns}$) \cite{NS}.
\begin{equation}
    P_{ns} = hf\left( \frac{1}{2} + \frac{1}{e^{\frac{hf}{k_BT}}-1}\right)
\end{equation}
Comparing the JPA-\emph{off} noise temperature ($T_n^{OFF}$), gain ($G^{OFF}$), and power spectrum ($P^{OFF}$) to the JPA-\emph{on} gain ($G^{ON}$), and power spectrum ($P^{ON}$), we estimate the noise temperature ($T_n^{ON}$) of the chain with JPA \emph{on}
\begin{equation}
    T_{n}^{ON} = \frac{1}{k_{B}B}\frac{G^{OFF}}{G^{ON}}\frac{P^{ON}}{P^{OFF}}(k_BBT_{n}^{OFF}+P_{ns}) - P_{ns},
\label{equ2}
\end{equation}
where \textit{B} is the bandwidth.
The system noise is the sum of $P_{ns}/k_BB$ from the NS and added noise $T_{n}^{ON}$ from the chain. The measurement results are shown in Fig.~\ref{gory}.
\section{Conclusion}
The RF chain with three JPAs as the first amplifier called Gorynych design has been experimentally studied. The chain is dedicated for axion haloscope experiments. Three separate resonances corresponding to each JPAs were observed. JPA-A has shown a periodic dependence on the external magnetic field and has shown amplification with the noise temperature lower than \SI{110}{\milli\kelvin}. We have characterized the chain with a JPA added noise between \SI{65}{\milli\kelvin} and \SI{110}{\milli\kelvin} at the frequencies between \SI{0.938}{\giga \hertz} and \SI{0.963}{\giga \hertz}.
\section*{Acknowledgement}
This work is supported in part by the Institute for Basic Science (IBS-R017-D1) and JST ERATO (Grant No. JPMJER1601). Arjan F. van Loo was supported by a JSPS postdoctoral fellowship.

\end{document}